%
%

\input harvmac.tex

\def\IR{\relax{\rm I\kern-.18em R}}
\def\IZ{\relax\ifmmode\mathchoice
{\hbox{\cmss Z\kern-.4em Z}}{\hbox{\cmss Z\kern-.4em Z}} 
{\lower.9pt\hbox{\cmsss Z\kern-.4em Z}}
{\lower1.2pt\hbox{\cmsss Z\kern-.4em Z}}\else{\cmss Z\kern-.4em
Z}\fi}
\font\cmss=cmss10 \font\cmsss=cmss10 at 7pt

\def\dd{{\cal D}}

\lref\juan{J. Maldacena, ``The Large N Limit of Superconformal Field 
Theories and Supergravity'', hep-th/9711200.}
\lref\klebanov{S.S. Gubser, I.R. Klebanov and A.W. Peet, 
``Entropy and Temperature of black 3-branes'', hep-th/9602135; 
S.S. Gubser and I.R. Klebanov, 
``Absorption by Branes and Schwinger Terms in the World Volume Theory'', 
hep-th/9708005.}
\lref\wittena{E. Witten, ``Anti de Sitter Space and Holography'', 
hep-th/9802150.}
\lref\igor{S.S. Gubser, I. Klebanov and A. Polyakov, 
``Gauge Theories Correlators from Non-Critical String Theory'', 
hep-th/9802190.}
\lref\polyakov{A.M. Polyakov, ``String Theory and Quark Confinement'',
hep-th/9711002.}
\lref\martinec{E.J. Martinec, ``Matrix Models of AdS Gravity'', 
hep-th/9804111.}
\lref\cone{
G. Moore and N. Seiberg, ``From Loops to Fields in 2-D Quantum Gravity'',
Int. J. Mod. Phys. A7,2601,1992; 
``Two Dimensional Quantum Gravity and Random Surfaces'', 
D.J. Gross, T. Piran and S. Weinberg Eds. World Scientific, 1992 
and references therein;
M. Natsuume and J. Polchinski,
``Gravitational Scattering in the C=1 Matrix Model'', hep-th/9402156.}
\lref\natidine{M. Dine and N. Seiberg, 
``Comments on Higher Derivatives Operators in Some SUSY Field Theories'',
hep-th/9705057.}
\lref\brsda{M. Bershadsky, Z. Kakushadze and C. Vafa, 
``String Expansion as a Large N Expansion of Gauge Theories'',
he-th/9803076.}
\lref\brsdb{M. Bershadsky and A. Johansen, 
``Large N Limit of Orbifold Field Theories'', hep-th/9803249} 
\lref\dealwis{S.P. de Alwis, 
``Supergravity, the DBI Action and Black Hole Physics'',
hep-th/9804019.} 
\lref\threeb{M.B. Green and M. Gutperle, ``Comments on Three-Branes'',
 hep-th/9602077.}
\lref\gsw{M.B. Green, J.H. Schwarz and E. Witten,
``Superstring Theory'', Vol 2., Cambridge University Press, 1987.}
\lref\schwrzb{J.H. Schwarz, 
``Covariant Field Equations of N=2, D=10 Supergravity'',
Nucl. Phys. B226, 269, 1983.}
\lref\witsus{L. Susskind and E. Witten, 
``The Holographic Bound in Anti-De-Sitter Space'', hep-th/9805114.}
\lref\matbh{T. Banks, W. Fischler, I.R. Klebanov and L. Susskind, 
``Schwarzschild Black Holes from Matrix Theory'', hep-th/9709091;
I. Klebanov and L. Susskind, ``Schwarzschild Black Holes in Various
Dimensions'', hep-th/9709108; E. Halyo, ``Six Dimensional Matrix Black
Holes in M(atrix) Theory, hep-th/9709225.}
\lref\matbhb{ 
G. Horowitz and E. Martinec, 
``Comments on Black Holes in Matrix Theory'', hep-th/9710217;
M. Li, ``Matrix Schwarzschild Black Holes in the Large N Limit'',
hep-th/9710226;
T. Banks, W. Fischler, I.R. Klebanov and L. Susskind, 
``Schwarzschild Black Holes from Matrix Theory II'', hep-th/9711005;
H. Liu and A.A. Tseytlin, 
``Statistical Mechanics of D0-branes Black Hole Thermodynamics'',
hep-th/9712063;
T. Banks, W. Fischler and I. Klebanov, 
``Evaporation of Schwarzschild Black Holes in Matrix Theory'', 
hep-th/9712236;
N. Ohta and J.-G Zhou, 
``Euclidean Path Integral, D0-branes and Schwarzschild Black Holes in 
Matrix Theory'', hep-th/9801023;
M. Li and E. Martinec, ``Probing Matrix Black Holes'', hep-th/9801070;
D. Kabat and G. Lifschytz, ``Tachyons and Black Holes Horizons in 
Gauge Theories'', hep-th/9806214.}
\lref\ansatz{A. Kehagias, 
``New Type IIB Vacua and their F-Theory Interpretation'', hep-th/9805131;
O. Aharony, A. Fayyazuddin and J. Maldacena, 
``The Large N Limit of N=2,1 Field Theories from Threebranes in 
F-Theory'', hep-th/9806159.}
\lref\watidan{D. Kabat and W. Taylor, 
``Linearized Supergravity from Matrix Theory'', hep-th/9712185.}
\lref\david{D. Lowe, 
``Statistical Origin of Black Hole Entropy'', hep-th/9802173.}
\lref\das{S.R. Das and S.P. Trivedi, ``Three-brane Action and the 
Correspondence between N=4 Yang-Mills Theory and $AdS$ Space'', 
hep-th/9804149;
S. Ferrara, M.A. Lledo and A. Zaffaroni, 
``Born-Infeld Correlations to D3-brane Action in $AdS(5)\times S(5)$ 
and N=4, D=4 Primary Superfields'', hep-th/9805082.} 
\lref\hight{G.T. Horowitz and S.F. Ross, ``Possible Resolution of 
Black Hole Singularities from Large N Gauge Theory'', hep-th/9803085;
E. Witten, ``Anti-De-Sitter Space, Thermal Phase Transition, and
Confinement in Gauge Theories'', hep-th/9803131; D.J. Gross and
H. Ooguri, ``Aspects of Large N Gauge Theoey Dynamics as Seen by
String Theory'', hep-th/9805129.}
\lref\mikewati{M.R. Douglas and W. Taylor IV, ``Branes in the Bulk of 
Anti-de-Sitter Space'', hep-th/9807225.}



\Title{\vbox{\baselineskip12pt\hbox{hep-th/9807230}
\hbox{IASSNS-HEP-98/69}}}
{\vbox{\centerline{A Proposal on Some Microscopic Aspects}
\centerline{}
\centerline{of the AdS/CFT Duality}
}}

\centerline{Micha Berkooz} 
\smallskip
\centerline{Institute for Advanced Study}
\centerline{Princeton, NJ 08540, USA}
\centerline{berkooz@sns.ias.edu}

\smallskip

\bigskip
\bigskip
\noindent

\centerline{Abstract}

\noindent We suggest a model of the large N limit ${\cal N}=4$ $D=4$ 
$SU(N)$ SYM as a gas of 3-branes in a 10 dimensional space. Field
theory analysis suggests that this 10 dimensional space does not carry
the usual gravity dynamics but rather a contraction of it. Using a
non-local transformation some aspects of the dynamics of this system
are mapped to the dynamics of standard gravitons on $AdS_5\times
S^5$. In particular some of the correspondence between operator in the
CFT and states on $AdS$ is more transparent.

\Date{July 1998}


\newsec{Introduction}

A recurring problem in black hole physics is the following. Suppose we
have a particle falling towards a cluster of D-branes which form a
black hole, then this process has 2 dual descriptions. The first is in
terms of the particle as an object in the field theory of the
branes, and the other is in terms of a particle moving in the well
defined near horizon geometry of the black hole. The puzzle is to
directly relate these two descriptions.

The understanding of this and related issues for the case of D3-branes
has been the subject of intensive research, \klebanov, and was
crystallized in \juan\ where it was suggested that the large N limit of
${\cal N}=4$ U(N) SYM in 3+1 dimensions is equivalent to type IIB
gravity on $AdS_5\times S^5$. A quantitative procedure of extracting CFT
dimensions from the analysis of this sugra vacuum is given in
\wittena\ and \igor.

One way of trying to understand the microscopic aspects of the
$AdS/CFT$ duality is along the lines of \polyakov. We will take
another approach in which when $N\rightarrow\infty$ we will try and
pass to a collective coordinates description of the gauge theory
system (almost, we will be more precise about this below) as a gas of
3-branes fluctuating in a 10 dimensional target space. This 10
dimensional space is not the same as $AdS_5\times S^5$, rather it is
related to it by a non-local transformation. In this respect the
construction is reminiscent of $2d$ gravity constructions \cone\martinec.

Suppose we start with the theory on a sphere and $N$ is small. We can
try and understand the structure of the vacuum in the following
way. We would like to separate the variables in the path integral into
slow and fast variables and integrate over the fast ones. When the
number of branes is small one can suggest the following
separation. The slow variables are N copies of a $U(1)$ theories
(along what would like to be the flat directions\foot{It is important
to emphasize that we are not working along real flat directions. These
are lifted by the $RX^2$ term}) and the fast variables are the $W$
bosons. The (Euclidean) action contains \wittena 
\eqn\eucact{\int d^4x \bigl( g^{\mu\nu} Tr(\partial_\mu 
X^\alpha\partial_\nu X^\alpha)+ R(g) Tr(X^2)\bigr)} where the indices
$\mu$ and $\nu$ refer to coordinates on the sphere, $X$ are the 6
adjoint scalar fields, the index $\alpha$ runs from 1 to 6, and $R(g)$
is the curvature of the metric. Since the flat directions have been
lifted the branes are confined to a finite region, $B$, in the space
of eigenvalues of $X$. Since the number of branes is small then most
of the measure in the path integral is in configurations in which the
branes are separated (but still within $B$) and we can therefore use a
gas of brane approximation. When we integrate out the $W$'s, the fast
variables, we generate a long range force between the separated
branes. The singularity (when branes meet) in the this long range
force is not a conceptual problem in our case since for most of the
time the branes are separated and do not feel this singularity. 

In the following we will try and extend this picture to a large number
of branes. In this picture the dominant configurations in the field
theoretic path integral will be approximated by a density of 3-branes
in a 10 dimensional space made out of 4 $x$ coordinates and 6
eigenvalues of $X^\alpha$, $\alpha=1..6$. We will call this space the
$\lambda-x$ space. Treating the singularity in the effective
action when branes intersect will then be crucial, and the burden of
the proof is now shifted to controlling the corrections to the
effective Lagrangian and to the dynamics of the system. We do not
claim to have complete control over these, but we can obtain some
level of qualitative understanding of some aspects of the $AdS/CFT$
duality.

Similar ideas have been proven useful in the discussion of Matrix
black holes \matbh\matbhb. In particular, references \matbhb\ discuss
the description of Schwarzschild black holes using an effective
description in terms of a gas of 0-branes\foot{or that a gas of
0-branes is the dominant component in the black hole.}. One important
distinction, however, should be made. Whereas most of these papers
discuss a form of mean field approximation of the gas, we will try and
analyze a large enough class of configurations in the path integral,
since we are in the vacuum of the theory and not in a high temperature
semiclassical regime. Nevertheless, for the lack of a better name, we
shall still refer to the picture as that of a gas of branes.

The $\lambda-x$ space, in which the 3-branes fluctuate, can then be
related to $AdS_5\times S^5$. This is done in a way that reproduces
some of the rules of associating operators in the field theory with
states in the bulk \wittena\igor. One should emphasize, however, that
two space are not identical. Rather, the transformation between
them is non-local (as in \cone\martinec). This makes precise the idea
that the branes ``are everywhere in $AdS$''.

The framework suggested above seems to have the following
implication. Within the gas of branes picture one expects to find a
well behaved static supergravity description of supersymmetric field
theories only if the gas of branes has a well defined stationary
state. In order for the gas to have a ground state we need to lift the
flat directions, or at least not allow, by any other means, for the
cloud of branes to disperse. In our case this is done by the
$R\lambda^2$ term but it can also be done using high temperature. Both
such configurations have well behaved supergravity descriptions
\wittena\hight\ and this is to be contrasted with situations in which
the gas is allowed to dissolve such as the system of 0-branes where
the supergravity solution is singular.

The organization of this paper is the following. Section 2 contains
some scaling arguments in $AdS$ that hint towards the interpretation
as a gas of branes. Section 3 describes some aspects of the non-local
transformation from $R^4\times R^6$ to $AdS_5\times S^5$. Section 4
sets up part of the machinery in the context of the free theory,
before we include the effects of the W bosons. Section 5 describes
corrections to the free theory, and the introduction of what will
become the supergravity fields.  Section 6 describes the
transformation of these fields to the $AdS$ and the matching between
some operators and excitations on $AdS$. Section 7 contains some
conclusion and open problems.

One important point of notation that we will use extensively is the
following. When we will think of the 6 scalars as $N\times N$ matrices
we will denote them by $X$. When we will think about 6N scalar fields
which are the N eigenvalues in the gas of branes picture, we will
denote these fields by $\lambda(x)$. When we will consider these 6N
scalar fields as embedding, for every point $x$, into an additional 6D
space then we will also use $\lambda$ to denote coordinates on this
space. In this case we will denote use the notation $\lambda$ without
any $x$ dependence.

As this paper was prepared for submission, related work appeared in 
\mikewati.

\newsec{The UV-IR relation}

Let us briefly review the relation \witsus\ between the IR regulator
of $AdS$ and the UV cutoff of the field theory on the boundary. In the
coordinates in which $AdS_5$ is
$${R^2\over (1-z^2)^2} {\displaystyle\sum_{i=1}^5{dz^i}^2}$$ one
imposes a cutoff $\sum {z^i}^2=1-\delta$. This is an IR cutoff on
the $AdS$ and a $UV$ cutoff on the field theory which lives at
$z^2=1-\delta$.

Another coordinate system, which will be more convenient for our
purposes, is the one in which the metric takes the form (neglecting
numerical coefficients)\juan
\eqn\mtrc{ds^2=\alpha'\biggl({1\over\sqrt{g^2N}}U^2dy^2+
\sqrt{g^2N}{dU^2\over U^2}\biggr)}
where $g$ denotes $g_{YM}$.  The IR-UV relation in \witsus\ can be
 transferred to the description in terms of these coordinates. In this
 case we fix $U_0$ as our sugra IR regulator, and this defines a UV
 regulator, $L_{uv}$, in the field theory. The relation that one
 obtains is
\eqn\scala{U_o L_{uv}=\sqrt{g^2N}.} This relation is given by the 
following analysis. If we define $U=\sqrt{g^2N} {\hat U}$ then the
metric is such that all the $N$ dependence is in a factor in front of
the metric, and the rest is ${\hat U}^2dy^2+{d{\hat U}^2\over {\hat
U}^2}$. Since the computation in \witsus\ is that of geodesics in the
metric then the UV-IR relation is ${\hat U}_0L_{uv}=1$, which is
\scala.

Loosely, we can think of the coordinate $U$ as some characteristic
size or scale \juan\ associated with the scalar fields on the brane,
which usually parameterize its position. One therefore would like to
know what is the significance, in the regulated field theory, of $U_o$
as a value of the scalar fields. The normalization that we will use is
such that the action (for a single brane) is
$${\cal L}\sim {1\over g^2}(\partial X)^2$$ where $X$ denotes the
scalar fields.

The interpretation of this scale is the following. Let us regulate the
theory in the IR by some scale which we will take as 1 and in the UV
by a length cutoff which we will denote by $\delta$ ($\rightarrow
0$). This is the same as we had on the sphere (we will also neglect
the conformal coupling $RX^2$ since the zero-mode of $X$ will not
play an important role in what follows). Expanding in momentum modes
we obtain that the action for a $U(1)$'s worth of $X$ is
\eqn\euncp{{\cal L}\sim {1\over g^2}\sum_{n,l} 
n^2X_{n,l}^2} where the factor of $n^2$ comes from the kinetic
term. $n$ labels the total momentum of the mode and $l$ parameterizes
the states in that momentum shell ($l=1..d_n$).

If we now wish to evaluate the dispersion of the values of $X$,
we can compute the quantity $<X(x)^2>$. For a single field this
yields
\eqn\disp{<X(x)^2>=g^2\sum_{n=1}^{1\over\delta} d_n {1\over n^2}=}
$$=g^2\sum_{n=1}^{1\over\delta} {n^3\over n^2}\propto g^2{1\over
\delta^2}$$ which is the same (up to constants, and for $N=1$) as the
relation \scala\ between $U_o$ and $\delta$.

For a large number of branes we have an $N\times N$ matrix. If we take
one of the $X$ matrices then we can diagonalize it. When we do so the
distribution of eigenvalues scales like $\sqrt{N}$. In our
normalization, where for a large number of branes ${\cal L}\sim \int
Tr(\partial X^2)$, the distribution of eigenvalues is governed by the
scale ${1\over N}Tr{X^2}\sim {N\over\delta^2}$ which sets a scale
${1\over\delta}\sqrt{N}$.

This scaling relation supports the frame work that was suggested
before. One may be able to think of the region in $AdS$ inside the
cutoff $\delta$ as corresponding to the region of eigenvalues in which
the field theory scalars fluctuate\foot{In fairness we should note
that there might be another scaling. We may argue that each of the N
branes is positioned in $\lambda$ according to the expectation value
of the scalar field. In this case the dispersion of the branes in the
eigenvalue space is much smaller, and should only be calculated by the
zero mode. We believe that this is not the correct scaling because
there is no energetic reason to impose first the restriction to the
zero mode whereas the expression \euncp\ above is more accurate from
that respect. We will also see below that we do not use the
expectation values, rather the analysis will be in each value of $x$
independently. A more serious problem is that our estimate for the
actual dispersion is not really correct on configurations that
dominate the path integral due to correlations, the commutator term in
the action or because of quantum correction. However, it is not clear
how to estimate this effect.}. The sugra excitations should then be
explained as objects living on this range in the space of
eigenvalues. The rest of the paper is a speculative step in this
direction.

\newsec{The transformation from $\lambda-x$ space to $AdS_5\times S^5$}

\subsec{The $\lambda-x$ space and the transformation from $R^4$}

Even though it is important to have the term $R\lambda^2$ in order to
have a well defined supergravity dual, we can choose to work for most
purposes with the metric of $R^4$, which is what we will do from now
on. 

In the gas of brane picture we have $N$ embeddings of D3-branes into a
10 dimensional space. This space is given by 4 coordinates $x^\mu$
along the brane and by 6 coordinates $\lambda^\alpha$ which are
associated with 6 scalar fields. This seems to be the usual IIB
D3-brane in flat space picture but this is not quite so. The important
difference is that the $\lambda$ coordinates parameterize the flat
directions of the field theory and are distinct from the 4 $x$
coordinates, and therefore the dynamics in the different direction can
be significantly different. In particular we will not obtain standard
gravitational dynamics on the $\lambda-x$ space. Rather we would like
to argue that some aspects of the dynamics on the $\lambda-x$ space
will be equivalent to dynamics on $AdS_5\times S^5$ after a non-local
transformation. We will discuss the dynamics on the $\lambda-x$ space at
greater length later, and for now we will and construct a map from
$\lambda-x$ space to $AdS_5\times S^5$.

In this subsection we will discuss the transformation law of scalar
functions on $R^4\times R$ (the distinct $R$ component is a single
$\lambda$ coordinate) to functions on $AdS_5$. We will later restore
the full $S^5$. $R\times R^4$ will be parameterized by a single
$\lambda$ and four $x$, and $AdS_5$ will be parameterized by $y$ and
$U$ as in equation \mtrc.

The transformation that we will look for is the most general linear
transformation of the form
$${\tilde f}(u,y)=\int d\lambda d^4x f(\lambda,x) K(\lambda,u,x,y)$$
such that it intertwines the action of the conformal algebra on both
sides.

Since we know the transformation laws of $x$ and $\lambda(x)$,
we know the geometric action of the conformal group on the space
parameterized by $x$ and $\lambda$. We also know the action of the
conformal group on $U$ and $y$. Taking care of rotational and
translational symmetries in $R^4$ is an easy matter, and the
non-trivial restrictions come from $D$ and $K^\rho$ (the special
conformal transformations). The action of these generators on
functions $f(\lambda,x)$ and ${\tilde f}(U,y)$ is given by\foot{One
immediately sees that the $\lambda-w$ plane can not be simply
identified with the $\rho,v$ space since they transform differently
under conformal transformation.}
\eqn\gena{Df=(\lambda \partial_\lambda - x^\mu\partial_{x^\mu})f}
$${\tilde D}{\tilde f}=(U\partial_U-y^\mu\partial_{y^\mu}){\tilde f}$$
\eqn\genb{K^\rho f=\bigl(-2x^\rho\lambda \partial_\lambda + 
(2x^\rho x^\mu -\delta^{\rho\mu}x^2)\partial_{x^\mu}\bigr)f}
$$ {\tilde K}^\rho {\tilde f}=\bigl(-2y^\rho U \partial_U + 
(2y^\rho y^\mu -\delta^{\rho\mu}(y^2+{1\over U^2})\partial_{y^\mu}\bigr)
{\tilde f}.$$

Once we have identified the action of these generators, we can require
that the map intertwines them. Using \gena\ and \genb\
This fixes the form of the transform to be
\eqn\trnspln{{\tilde f}(u,y)=\int d\lambda d^4x\lambda^3
f(\lambda,x)K\biggl(\lambda u\bigl((x-y)^2+{1\over u^2}\bigr)\biggr)}
 
We are able to determine the transform up to a function of a single
variable. This is the best we expect to do if we use only the
invariance under the conformal group because of the following reason.
Functions on $AdS$ can be decomposed according to eigenvalues under
the $AdS$ Laplacian, which we will denote by $\Delta_{AdS}$, and this
value does not change under the action of conformal
transformations. On the $\lambda-x$ there will be another differential
operator, which we will also refer to as a ``Laplacian'' (although
there is no metric on that space), and denote by $\Delta_\lambda$
which plays a similar role. The transform maps these two operators to
each other, i.e., it maps a function with a given eigenvalue under
$\Delta_{AdS}$ to a function with the same eigenvalue under
$\Delta_\lambda$, but we can allow the transform to map it to such a
function times an arbitrary constant. An independent multiplication is
allowed for every value of the eigenvalue under the $\Delta$ operators
and hence the transform is determined up to a function of a single
variable.

\subsec{The ``Laplacian'' on the $\lambda-x$ plane}

The ``Laplacian'', $\Delta_\lambda$, on the $\lambda-x$ side will play
an important role in what follows. Returning to 6 $\lambda$'s, it turns
out that this operator is
\eqn\laplmbd{\Delta_\lambda = (\sum_\beta {\lambda^\beta}^2)
\sum_\alpha {\partial^2\over\partial{\lambda^\alpha}^2}}
 
The computation is the following. The Laplacian on the $U,y$
coordinates is
$${1\over U^3}\partial_u U^5\partial_u +{1 \over U^2}\partial_y^2.$$
We can transform it to the $\lambda-x$ plane using the kernel above,
and obtain that
$$\int d\lambda d^4x \lambda^3\phi(\lambda,x) 
\biggl({1\over U^3}\partial_U U^5\partial_U+
                                       {1\over U^2}\partial_x^2\biggr)
K\biggl(\lambda u \bigl(t^2+{R^4\over u^2}\bigr)\biggr)=$$ 
$$=\int d\lambda d^4x \lambda^3\phi(\lambda,x)
(\lambda^2\partial_\lambda^2+5\lambda\partial_\lambda)K=$$
$$=\int d\lambda d^4x \lambda^5 
((\partial_\lambda^2+{5\over\lambda}\partial_\lambda)\phi(\lambda,x))K.$$

We are now interested in restoring the 6 $\lambda$ instead of having
only one. To do so we need to add an $S^5$ angular part to the kernel,
which we will not discuss in great detail. More interesting is the
effect on $\Delta_\lambda$. Choosing an eigenvalue $\omega^2$ for the 
$S^5$ Laplacian, the operator 
${1\over U^3}\partial_U U^5\partial_U + {1\over U^2}\partial_y^2+w^2$ 
is now mapped to 
$$=\int {d^6\lambda d^4x\over \lambda^2} 
\lambda^2\biggl(\bigl(\partial_\lambda^2+{5\over\lambda}\partial_\lambda+
{w^2\over\lambda^2}\bigr)\phi(\lambda,x)\biggr)K,$$ where we have
separated the scale invariant measure from the operator.  One now
recognizes the operator as ${\lambda^\beta}^2{\partial^2\over\partial
{\lambda^\alpha}^2}$. 	A more group theoretic approach that yields the
same result is to realize that the $\Delta$ operators are the Casimir
operators for the conformal group. This allows their immediate
computation in both spaces.

\subsec{Locality on the boundary}

Even though we do not have an explicit form for the kernel of the
transformation one can see that it is local on the boundary for
physical states in the sense that if we take a function that satisfies
the Laplacians on both sides and is supported at a given $x$ then it
will be supported at $y=x$ as $u\rightarrow\infty$. This can be
derived by using only group theory properties of the transformation.

On the $\lambda-x$ side we will choose a function of the form
$P(\lambda)\delta(x)$ where $P(\lambda)$ is a symmetric traceless
polynomial, $\Delta_\lambda P(\lambda)=0$. This function is also
annihilated by $K_\mu$. Since the transformation commutes with
$SO(5,1)$, the same will be true for image of the function under the
transformation. These gives us two equations (taking into account the
$SO(6)$ quantum numbers, which determines the mass $m$, and $SO(4)$
quantum numbers)
$$(\Delta_{AdS}+m^2){\tilde f}(U,y)=0$$
$$K^\mu_{AdS}{\tilde f}=-2y^\mu U\partial_U +\biggl(2y^\mu
y^\nu-\delta^{\mu\nu}\bigl(y^2+{1\over U^2}\bigr)\biggr)
\partial_\nu {\tilde
f}(U,y)=0.$$ The solutions (for every $m$) to these equation are the
propagators from the boundary to the bulk described in \wittena\ and
\igor. These functions are such that as we go to the boundary the 
support of the function is at $y\rightarrow x$.

\subsec{A brief look at whats ahead}

The axion-dilaton pair are scalar fields on ${\cal M}=AdS_5\times
S^5$, and their linearized equation of motion is $\Delta_{\cal
M}\Phi=0$. If we would like to describe similar fields on the
$\lambda-x$ space their equation of motion will be 
$${\partial^2 \over \partial
\lambda^2}\Phi(\lambda^\alpha,x^\mu).$$ Note that the field is 10
dimensional but its equations of motion are 6 dimensional - only
derivatives in the $\lambda$ directions appears. Since the $\lambda$
space is the space of flat directions we expect such a result. As we
will see later the field $\Phi$ participates in the mediation of long
range forces in the $\lambda-x$ space. Since these long range forces
are supposed to generate an overall field theoretic effect, they
should have a correct expansion in terms of series of local (in $x$)
terms. This implies that the long range forces should be such that
they connect branes at different point in $\lambda$ but at the same
value in $x$. This is exactly what the 6D equation of motion achieves.

This transformation also improves the ``locality'' of the
description. We see that using the transform we can make a function
that is spread in $y$ (except in the boundary) into a function that is
localized in $x$. This will be the key to obtaining, after we have
laboriously crunched through all the gas description and several
additional approximations, the correspondence between spread states on
$AdS$ and local operators in the field theory.

\newsec{The free ``Gas'' of branes}

The picture that we will describe is that of N branes fluctuating in
the $\lambda-x$ space. For each brane we have the fields
$\lambda_i^\alpha(x)$, $A_\mu^i(x)$ and fermions, where $i=1..N$,
$\alpha=1..6$ and $x$ are four dimensional coordinates. When writing
down the action we will focus on the bosonic part.

The lowest order action ``along the flat direction'' is 
\eqn\acta{\sum_{i=1}^N\int d^4x \biggl(
\partial^\mu \lambda_i^\alpha(x)\partial_\mu \lambda_i^\alpha(x)
+{1\over g_0^2}F_i^2\ (+\ R\lambda(x)^2) \biggr) } (The term in the
brackets depends on whether we are on the sphere or not. For the
most part we will work on $R^4$).

Had these been a large number of particles moving in a confined
region, we would have known what to do. The correct prescription would
be to go to collective coordinates of this gas, i.e. to variables such
as local density and local average velocity. A generalization of these
quantities appears in this problem as well. For example
$\partial_\mu\lambda^\alpha$ is a generalization of
velocity. Furthermore, since we have additional degrees of freedom in
the field theory we will have additional collective degrees of
freedom in the $\lambda-x$ space, such as $A_\mu(\lambda,x)$. The main
assumption that we need to make is that when the branes are close
together they fluctuate roughly the same such that it is useful to
describe them using collective excitations. 

The quantities that we will find useful to work with are
$L_\mu^\alpha$ which is roughly $\partial_\mu X^\alpha$, $A_\mu$ and
$\psi(\lambda,x)$. The function $L_\mu^\alpha(\lambda,x)$ measures the
following: Suppose we are in a point $\lambda_0$ and $x_0$. There is a
3-brane that passes through this point. Let us denote the trajectory
of this brane by $U^\alpha(x;
\lambda_0,x_0)$ and we will then define
$L_\mu^\alpha(x_0,\lambda_0)=\partial_{x^\mu} U^\alpha(x_0;
\lambda_0,x_0)$, so the L-matrix measures the gradients of the branes at
a given point. From it we can obtain $\lambda^\alpha(x)$ of the brane
everywhere but this is a non-local transformation and we will avoid
using it. If the only $\mu$ coordinate was the time coordinate then
this quantity would have the velocity profile of the gas. Since we
require that $L^\alpha_\mu$ has integral surfaces, then not all the
components of $L$ are independent. This is take care of below by the
Lagrange multiplier $M^{\mu\nu}_\alpha$.

The definitions for $A_\mu(\lambda,x)$ and $\psi(\lambda,x)$ are
simpler. If brane $k$ passes through point $(x_0,\lambda_0)$ then we
will define $A_\mu(x_0,\lambda_0)=A_\mu^k(x_0)$, and a similar
definition holds for the fermions.

Another object that will be useful is a derivative along the
brane. This is given by 
\eqn\defd{{\cal D}_\mu = {\partial\over\partial {x^\mu}}+
L^\alpha_\mu {\partial\over\partial\lambda^\mu}.}  This expression is
obtained when lifting $\partial\over\partial x^\mu$ to the
$\lambda-x$ space.

In order to write the action we need one more ingredient which is the
density of branes in each point in the $\lambda-x$ space. We will
denote this function by $\rho(\lambda,x)$. It is a density only on the
$\lambda$ coordinate and is normalized to 1 at each $x$
separately
$$\int d^6\lambda \rho(\lambda,x)=1.$$ This function is not a new
independent variable since it depends on $L_\mu^\alpha$ (although not
completely determined by it). If we start from a point
$(\lambda_0,x_0)$ and follow the integral surface determined by $L$
then we can deduce how the density of branes changes along such
surfaces. Since this information is encoded both in $\rho$ and in
$L$, there is some redundancy between these quantities. The redundancy
will be of the form that the variation of $\rho$ as we change $x$ and
go along integral lines of $L$ should match a change in a volume
element which is determined by $L$. This constraint is taken care of
by the Lagrange multiplier $M^\mu$.

The action that one obtains is (up to numerical coefficients) 
\eqn\acttree{N\int d^6\lambda d^4x \rho(\lambda,x)
\biggl(L^\alpha_\mu L^\alpha_\mu+
{1\over g_0^2} F_{\mu\nu}+
i{\bar \psi}\sum\partial \psi\biggr)+}
$$\int d^6\lambda d^4x \biggl( M^{\mu\nu}_\alpha \partial_\mu
L_\nu^\alpha+ M^\mu\bigl(\partial_\mu(\rho)+\partial_\alpha(\rho
L^\alpha_\mu\bigr)\biggr) ).$$

\subsec{Supersymmetry}

Since this system is derived from the system of N 3-branes it
possesses the full superconformal symmetry. We will, however, make
this slightly more explicit. The path integral that we now do is over
configurations of N branes fluctuating in the $\lambda-x$ space and we
would like to show how this configuration transforms under some of the
superconformal transformation.

In this section we will briefly discuss the some aspects 16
supersymmetries of the model. Some aspects of the conformal symmetry
will be discussed in the next subsection. Overall they will generate
the full superconformal algebra.

The supersymmetries of the model are easy to guess. The only change
one needs to do is to replace $\partial_\mu$ by $\dd_\mu$ from
\defd. An N=1 worth of supersymmetries can be described using a 
superspace formalism. As usual one adds the usual fermionic
coordinates $\theta$ and $\bar\theta$. The only difference is in the
definition of the operators $Q,{\bar Q},D,{\bar D}$. These are defined
such that ${\dd}_\mu$ replaces $\partial_\mu$.

There are two features that one needs to preserve in order to use the
conventional superspace construction of N=1 theories. The first is
that
$$[{\dd}_\mu,{\dd}_\nu]=0,$$ and this is true after we solve the
constraint on $L_\mu^\alpha$ (implemented using a Lagrange multiplier)
which reconstructs the embedding of the brane into $\lambda$ space.
The 2nd property is the ability to integrate by part. This is also true
in our case, i.e., 
$$\int d^4xd^6\lambda \rho(\lambda,x) f_1(\lambda,x){\dd}_\mu
f_2(\lambda,x)= -\int d^4xd^6\lambda \rho(\lambda,x) ({\dd}_\mu
f_1(\lambda,x)) f_2(\lambda,x)$$ because of the dependence between
$\rho$ and $L^\alpha_\mu$ (which is again implemented with a Lagrange
multiplier).

Using this superspace formulation, the appropriate definition of a
chiral superfield is by the condition
$${\dd}_\alpha \Phi(\lambda,x)=0.$$ The chiral fields that we will use
are the chiral fields whos lowest component is the coordinate
$\lambda^\alpha$ (made into 3 complex pairs) which we will by
$\Lambda^\alpha$. We will also have a $U(1)$ vector multiplet W (whos
chirality properties are also defined by $\dd$). 

The last ingredient that we need is the transformation law of
$\rho$. As we will see momentarily, these supersymmetry transformation
laws also include a flow term because $\lambda$ changes under the
supersymmetry. We require that $\rho$ transforms as a density under this
flow and this defines its susy transformation. The Lagrangian that one
obtains is
\eqn\sprspclag{\sim \int d^6\lambda d^4x \rho(\lambda,x) 
\biggl(\int d^4\theta \Lambda^{\alpha\dag} \Lambda^\alpha+
 \int d^2\theta W^2 + c.c. \biggr)}

The expansion of the chiral field $\Lambda^\alpha$ is the following
\eqn\chrlexnd{\Lambda^\alpha(\lambda,x)=
\lambda^\alpha+
\theta^a\psi_a^\alpha(\lambda,x)+
\theta\sigma^\mu{\bar \theta}L_\mu^\alpha+...}  
The supersymmetry variation of this multiplet (up to numerical
coefficients) is
\eqn\susyvarc{
(\zeta^aQ_a+{\bar\zeta}_{\dot a}Q^{\dot a})\Lambda^\alpha\sim
\Lambda^\alpha+\zeta^\alpha\psi^\alpha_a(\lambda,x) + 
\theta^a(\psi_a+\sigma^\mu_{a{\dot a}}{\bar\zeta}^{\dot a}L^\alpha_\mu)
+...}  The lowest component in the expansion signifies the position of
the field, therefore \susyvarc\ encodes the fact that we changed the
embedding of the D3 brane into $\lambda$ space when we do a susy
transformation. As functions on the $\lambda-x$ space the physical
degrees of freedom in the $\Lambda^\alpha$ multiplet trasnfrom as
\eqn\acmptrns{\delta L^\alpha_\mu\sim
-\bigl( (\zeta\psi^\beta){\partial\over\partial\lambda^\beta}+
   ({\bar{\zeta\psi}}^{\bar\beta})
         {\partial\over\partial{\bar\lambda}^{\bar\beta}}\bigr)
L^\alpha_\mu+ \zeta{\dd}_\mu\psi^\alpha}
$$\delta \psi^\alpha= -\bigl(
(\zeta\psi^\beta){\partial\over\partial\lambda^\beta}+
({\bar{\zeta\psi}}^{\bar\beta})
{\partial\over\partial{\bar\lambda}^{\bar\beta}}\bigr)\psi^\alpha+
+\sigma^\mu{\bar\zeta}L_\mu^\alpha$$ 
and a similar transformation, which we will not make explicit, for 
the vector multiplet.

\subsec{Conformal Symmetry}

We would like to show how the conformal symmetries act on fields in
the $\lambda-x$ space. The action of the conformal group has two
pieces to it. The first is a geometric action in the $\lambda-x$ space
and the 2nd is the standard dimension dependent rescaling. The
geometric action is the following:

\eqn\cnfgeom{P_\mu={\partial_{x^\mu}}}
$$M^{(g)}_{\mu\nu}=x^\mu\partial_{x^\nu}-x^\nu\partial_{x^\mu}$$
$$D^{(g)}=\lambda^\alpha\partial_{\lambda^\alpha}-x^\mu\partial_{x^mu}$$
$$K^{(g)}_\rho=-2x^\rho\lambda^\alpha\partial_{\lambda^\alpha}+
(x^\rho x^\nu-\delta^{\rho\nu}x^2)\partial_{x^\nu}$$

And after the inclusion of the dimension dependent rescaling, the
final action on a field $\Phi$ who's dimension is $d$ is
\eqn\cnffin{P_\mu\Phi(\lambda,x)={\partial_{x^\mu}}\Phi(\lambda,x)}
$$M_{\mu\nu}\Phi(\Lambda,x)=
\bigl(x^\mu\partial_{x^\nu}-x^\nu\partial_{x^\mu}+
\Sigma_{\mu\nu}\bigr)\Phi(\lambda,x)$$
$$D\Phi(\lambda,x)=(-d+\lambda^\alpha\partial_{\lambda^\alpha}-
x^\mu\partial_{x^mu})\Phi(\lambda,x)$$
$$K_\rho\Phi(\lambda,x)=
\bigl(-2x^\rho(\lambda^\alpha\partial_{\lambda^\alpha}-d)+
(x^\rho x^\nu-\delta^{\rho\nu}x^2)\partial_{x^\nu}-
x^\mu\sigma_{\mu\rho}\bigr)\Phi(\lambda,x)$$

These transformation rules may be derived as follows. Given a field
$\Phi_i(x)$ on $x$ with dimension $d$, its variation on the
$\lambda-x$ space can also be split into the following two parts. The
first, which we will denote by $\delta_0$, is just its 4D variation
pulled back to the $\lambda-x$ space, and the second is a transport
term that arises because we change $\lambda(x)$ which is the position
of the brane. The total variation is therefore
\eqn\ttlvar{\delta \Phi(\lambda,x)=\delta_0\Phi-
(\delta_0\lambda^\beta){\partial\over\partial_{\lambda^\beta}}\Phi}
where for scale transformation
$$\delta_0\Phi= (-x^\mu{\dd}_\mu+d)\Phi$$
$$\delta_0\lambda^\beta=-x^\mu L_\mu^\beta + \lambda^\beta$$
and for special conformal transformations
$$\delta_0\Phi= (2x^\rho x^\mu{\dd}_\mu-x^2{\dd}_\rho+dx^\rho)\Phi$$
$$\delta_0\lambda^\beta=2x^\rho x^\mu
L_\mu^\beta-x^2L_\rho^\beta+x^\rho \lambda^\beta.$$ Inserting these
into \ttlvar yields the expression above. Note that in $\delta_0$ one
needs to use $\dd_\mu$ and not $\partial_\mu$. The former is the
correct extension of a derivative along the brane when going to the
$\lambda-x$ picture.

In this notation $A_{\mu}$ has dimension 1, and the worldvolume
fermion has dimension ${3\over2}$. $\rho$ is a density and so
transforms in way that $d^6\lambda \rho(\lambda,x)$ has dimension 0
(so $\rho$ has dimension -6). Since $L_\mu^\alpha$ corresponds to
$\partial_\mu\lambda^\alpha(x)$ its transformation law is slightly
different but is still an immediate consequence of the field theory
transformation law.


Once we have demonstrated conformal invariance and supersymmetry we
obtain the entire superconformal symmetry.

\newsec{Factorizing the low-energy action}

We have seen that we can write down the action for the $N$ free D3
branes in the $\lambda-x$ space. However, in order to get a realistic
description of the dynamics one needs and take into account the
interactions of the branes along the flat direction due to the
W-bosons. Our approach is to start with the system at low density,
write the corrections in this regime and try and understand what
happens when we increase the density. After we have identified the
degrees of freedom at large density, we would like to write down an
effective action that controls these degrees of freedom.

Also, we will focus only on corrections along the flat directions
which are of the form $F^4$ or its susy partners. This is again very
similar to what is done in Matrix black holes. Since in ${\cal N}=4$
D=4 SYM this interaction is protected by supersymmetry, it will be the
most reliable to use in extrapolating between the different
regimes. These will actually suffice in matching some operators on the
boundary and excitations in the bulk.

\subsec{The $F^4$ term along the flat directions}

In this section we will briefly discuss some field theoretic aspects
of the $F^4$ term. First let us discuss the interaction between a pair
of branes. In this case the relevant degrees of freedom are a single
$U(1)$ multiplet on each of the two branes. The sum decouples, but the
difference is corrected at one loop, which is the infamous $F^4$ term
(and it superpartners). We will mainly in interested in the $F^4$. The
term that one obtains is \natidine
$${ (F_{+,1}-F_{+,2})^2 (F_{-,1}-F_{-,2})^2\over 
(<\lambda_1>-<\lambda_2>)^4 }$$
where $F_{+(-)}$ is the (anti)self-dual part of the field strength,
and the index $1,2$ is the index of the brane.

We are interested in the analogue of this term when the branes are
embedded in a more arbitrary way in the $\lambda-x$ space, i.e., when
the vev of each of the branes is varying (but st non-zero
separation). The full term under these circumstances is not known, but
we can guess some parts of it when the branes do not bend too much. In
that case we certainly expect that there will be a good expansion in
local terms. More precisely, since the original action was invariant
under Weyl rescaling (and up to C-number anomaly the same is true
quantum mechanically), we require the same of this term in the
effective action. Our conventions for the Weyl rescaling are
$$g_{\mu\nu}\rightarrow e^{2w(x)} g_{\mu\nu},\ X^\alpha(x)\rightarrow
e^{-w} X^\alpha(x),\ F(x)\rightarrow F(x).$$ Since we know how the
fields transform under Weyl rescaling and what the term is when the
vev's are flat, we obtain that at least part of the term for a varying
vev is of the form
\eqn\pairint{\int d^4x \sqrt{g}{(F_1(x)-F_2(x))^4 \over 
(<\lambda_1>(x)-<\lambda_2>(x))^4,}}
 i.e., the difference of the expectation value is evaluated at each
 point.

Another point that we need to understand is what happens to this term
when there are many branes. In this case there is no proven
non.-ren. theorem for the $F^4$ term, although one expects that the
term would still be restricted. As for the one loop contribution, in
this case it is the sum over the interaction between all
pairs. Suppose we have a loop with some external massless fields
attached to it (fields from the flat directions). All of these carry
charge 0 under the N $U(1)$'s that are unbroken along the flat
directions. The particles that we are integrating out can not change
their charge in the loop. Since they are charged under a pair of the
$U(1)$'s, say $k_1$ and $k_2$, the contribution of this diagram will
be the same as if we had only branes $k_1$ and $k_2$ which means that
it is included in the sum over all pairs.

In fact, the t'hooft double line notation lends itself to easy
manipulation along the flat directions. Since the external, particles
are associated with one of $N$ U(1) multiplets, the diagram becomes a
surface with several holes, each associated with an excitation of a
given brane (along the flat directions). This is quite clear from
open-closed string duality in the description of D-branes. A
diagramatic analysis along these lines is in fact expected to be much
simpler than usual (as for example in \brsda\brsdb) because the vertex
operator does not change the Chan-Paton index.

\subsec{Resolution of the $F^4$ singularity}

The singularity tells us that we have integrated out degrees of
freedom. These are the $W$ bosons. However, for our purposes we are
not very interested in the precise nature of these degrees of
freedom. Since we are working in a gas of branes approximation, a
generic pair of branes is separated from each other, the $F^4$ term is
not singular and the mass of the $W$'s is not zero. If most of the
effect of the correction term comes primarily from this regime, which
we will assume is the case, then any way that we choose to mimic this
term will give the same result\foot{For some processes at least} -
whether we choose to reintroduce the $W$'s, put it by hand (as is done
in many cases for Matrix black hole applications) or generate it by
other means. The rest of the construction relies on this freedom.

It is important to emphasize that we have made a very strong
assumption. As explained, this assumption is that the path integral is
dominated by configuration which are approximately a gas of separated
branes, and that the effect of the $W$'s that become light can be
encoded in an effective Lagrangian of these branes. It is difficult to
prove these assumption without a detailed dynamical analysis which we
do not know how to do. Rather we will assume it and examine whether we
can obtain any insight into the AdS/CFT correspondence under this
assumption. As we have mentioned before, this assumption has yielded
in the context of Matrix theory useful insight into Schwarzschild
black holes which are just as good as a classical sugra vacuum.

Other than the W-bosons, an exchange of a a sugra multiplet between
separated branes can also generate the $F^4$ term. By an ${\cal N}=4$
Non.-ren. theorem their contribution is the same. Our purpose is
therefore to introduce fields that are similar to the gravity
fields. Using these, one expects that we will be able to reproduce the
essential ingredients of the interaction between separate branes in a
similar fashion to the DBI action. These field, to which we will refer
as pseudo-gravity, live on the $\lambda-x$ space and will become the
10D sugra fields only after we perform the transformations described
in section 3.

More precisely what we will do is the following. For every
configuration in the path integral, i.e., a configuration of of the
D3-branes embedded in the $\lambda-x$ space, we will define a set of
fields, the pseudo-gravity fields, that satisfy a set of differential
equations with the fields on the branes ($L,\ F$ etc.) as sources. As
example of such an equation will be that of the ``dilaton-axion''
fields
\eqn\exmpl{\partial_\lambda^2\phi(\lambda,x)=
\rho(\lambda,x)F^2(\lambda,x)}
where $\phi$ is a pseudo-sugra field that will become a combination of
dilaton and axion after the transformation in section 3. These fields
are such that when we insert their value into the action we obtain the
$F^4$ correction to the action. The pseudo-gravity fields should be
thought of as auxiliary fields that satisfy their equations of motion
with the off-shell configuration of the 3-branes as a
source\foot{There might be an off-shell extension of the
pseudo-gravity fields but we will not need one.}. We would then like
to show, for a large enough class of configurations in the path
integral, how the matching between some operators and states in
$AdS/CFT$ correspondence comes about.

There might also be a description in which the pseudo-gravity
fields arise from dressing up collective excitations of the gauge
theory and in particularly of the $W$'s. The procedure that we will
discuss does so but only indirectly for most of the pseudo-sugra
multiplet. Returning to equation
\exmpl, if we are using only $\phi$ that satisfy this equation of
motion we could think of it as a definition of $\phi$ in terms of
variables $F^2$ and hence it is a object in the field theory. A similar
approach for the case of dynamics of D0 branes was taken in
\watidan\david.

We would like to emphasize again that the pseudo-sugra fields are NOT
10 dimensional sugra on the $\lambda-x$. There are several ways of
seeing this. One of the ways is to notice that we want the interaction
to be propagated only in $\lambda$, at constant $x$. This is different
from 10D sugra fields that propagate in all 10 dimensions. A related
reason is that in the vacuum of the theory we expect to retain the
conformal symmetry, but there is no conformally invariant metric in
the $\lambda-x$ space (in fact, we will never really use a metric
structure on the entire $x-\lambda$ space).

As we have explained above, and will make more explicit below, these
fields will become regular supergravity fields only after we use the
non-local transformation to map from the $\lambda-x$ space to the
$u-y$ space. Under such a transformation the non-sugra equations of
motion for the pseudo-gravity multiplet will be transformed into
supergravity equations of motion on $AdS_5\times S^5$ (at least, for
the cases that we have checked).

In that sense, in the analysis that we will present here there is
really no fundamental reason to go to the AdS, since the discussion is
consistent in the $\lambda-x$ space. It is therefore not clear which
of the two spaces is more fundamental. This question may be resolved 
by examining higher order interactions, or if we want to formulate
the theory as a string theory.

\subsec{Contraction of Gravity}

Since the fields that we will introduce will not satisfy exactly the
IIB supergravity equations of motion, and the gas of D3 branes is not
a gas of D3 branes in sugra, one needs to explain what equations they
do satisfy and how to couple them to the $U(1)^N$ field theory.

One would further like to couple them in a way that will generate the
$F^4$ term. Since the sugra couplings to D3 branes do reproduce this
term by tree level sugra exchange, we will take the usual sugra and
its D3 brane as a starting point and deform it until we obtain the
desired action for the pseudo-sugra fields. 

In this section we will discuss the behavior of the pseudo-sugra
fields. In the next one we will discuss its coupling to the brane. The
deformation that we will discuss is a sort of ``contraction'' of
gravity, and is closely related to going to the near horizon
geometry. 

Our initial starting point is that of D3-branes fluctuating in
$\lambda^\alpha-x^\mu$ space. In order to make this space into a 10D
space where all the coordinates are on equal footing we need to choose
a dimensionful parameter\foot{This is not the same as $l_p$ on
$AdS$. Since we will be dealing with the sugra equations of motion at
the linearized level (free, apart from sources) we will not be able to
determine what is $l_p$ on $AdS$.}, which we will denote by $l_p'$ and
define new coordinates
$$x^\alpha = \lambda^\alpha {{l'}_p}^2$$ where now all the coordinates
are of mass dimension $-1$.

To construct the pseudo-gravity multiplet on this space, we begin with
the 10D sugra multiplet and its equations of motion. Our coordinates
naturally split into a set of 4 coordinates ($x^\mu$) and another set
of 6 ($x^\alpha$). The equation of motion for the scalars field in
gravity is (assuming $g^{\mu\alpha}=0$)
\eqn\grvsclr{
 {1\over\sqrt{g}}\bigl(
  \partial_\alpha g^{\alpha\beta}\sqrt{g}\partial_\beta +
  \partial_\mu    g^{\mu\nu}     \sqrt{g}\partial_\nu
\bigr) \phi(x^\alpha,x^\mu)=0}
In order to obtain the equations of motion on the $x^\alpha-x^\mu$
space we will take
\eqn\grvcntrct{g^{\mu\nu}=\epsilon\rightarrow 0,\ 
               g^{\alpha\beta}=O(1).}  This is similar to taking the
limit of the near horizon geometry since in this limit the invariant
separation between two point in fixed $\lambda-x$ shrinks in the
$x^\alpha$ direction relative to the $x^mu$ directions.

When we do this scaling we would like to keep the N=4 SYM finite,
i.e., without any powers of $\epsilon$. In order to do so we can
identify the above rescaling with a constant Weyl rescaling in the
SYM. From this one deduces that the scaling of the SYM fields is the
following:
\eqn\symscl{F\sim O(1),\ L^\alpha_\mu\sim O(\epsilon^{1\over2}),\ 
\Psi\sim O(\epsilon^{3\over 4}),\ \eta\sim \epsilon^{-1\over 4}}
where $\eta$ is a parameter of one of the 16 susy (which are the only
ones that we will check in this section). One more requirement that we
will impose is that certain couplings of D3-brane fields to the sugra
fields remains finite as we take $\epsilon\rightarrow 0$. We will list
those when we will use them.

This information is enough to determine the scaling of all the
pseudo-sugra fields. This is done by requiring that there are no terms
in the deformed IIB susy transformations that are singular in
$\epsilon$. An example of such a computation is the following. Let us
denote by $\Gamma=\Gamma^{0123}$, which defines projection operators
$P_L={1\over 2}(1-i\Gamma),\ P_R={1\over2}(1+i\Gamma)$. The will take
the unbroken susy to be $\eta=\eta_R$.

The susy transformation laws in type IIB include the transformations
(for the complete transformation as well as other relevant conventions
see
\gsw\schwrzb):

\eqn\susyiiba{ \delta B_{\mu\nu}^i=
V_+^i{\bar\eta}^*
\Gamma_{{\bar\mu}{\bar\nu}}e^{\bar\nu}_{[\nu}e^{\bar\mu}_{\mu]}
(\lambda_R)^*+....}
\eqn\susyiibb{ \delta \lambda_R\sim
\Gamma^{{\bar\mu_1}{\bar\mu_2}{\bar\mu_3}}\eta
e^{\mu_1}_{\bar\mu_1}e^{\mu_2}_{\bar\mu_2}e^{\mu_3}_{\bar\mu_3}
G_{\mu_1\mu_2\mu_3}+...}  where $G_{\mu_1\mu_2\mu_3}\sim
V_i\partial_{[\mu_1}B^i_{\mu_2\mu_3]}$, and $i$ is an $SU(1,1)$ index
($B^i$ denotes the entire $SL(2,Z)$ multiplet of $B^{NSNS}$ and
$B^{RR}$).

The $V$'s are determined by the axion-dilaton scalars. Because we
would like to keep their leading coupling to the brane ($\int \phi
F^2$) and the operator to which they couple does not scale with
$\epsilon$, then the $\epsilon$-dimension of $V$ is zero. We have also
determined the scaling of the vielbein so the only unknown variables
in \susyiiba\ and \susyiibb\ are the scaling dimensions of
$B_{\mu\nu}$ and of $\lambda_R$. We will denote these by
$[B_{\mu\nu}]_\epsilon$ and $[\lambda_R]_\epsilon$.

The basic requirement is that there would be negative powers of
$\epsilon$ in the susy transformation. Otherwise the procedure will
not give a well defined end result. This requirement imposes the two
following inequalities.
$$\susyiiba \Rightarrow  [B_{\mu\nu}]_\epsilon \le 
-{5\over 4}+[\lambda_L]_\epsilon$$
$$\susyiibb \Rightarrow [\lambda_L]_\epsilon \le 
{5\over 4} +[B_{\mu\nu}]_\epsilon,$$ which gives an equality 
\eqn\exmpleql{[B_{\mu\nu}]_\epsilon = 
-{5\over 4}+[\lambda_L]_\epsilon.}  In this way one can obtain
equations between the dimensions of different fields in the sugra
multiplet. Another relation that involves $B_{\mu\nu}$ which we will
use comes from the requirement that the term $k^{1\over2} \int F\wedge
B$ has no $\epsilon$ dependence\foot{$k$ is Planck's constant to some
power. In our notation $F,B$ and $k$ has mass dimension 2, 4 and -4
respectively}. This gives a relation $${1\over
2}[k]_\epsilon+[B_{\mu\nu}]_\epsilon=0$$

The $\epsilon$-dimensions that one obtains using this procedure are:
\eqn\sclnga{[\Psi_\alpha]_\epsilon={9\over4},\ 
            [\Psi_\mu   ]_\epsilon={7\over4},\
            [\lambda    ]_\epsilon={9\over4},\
	    [k]_\epsilon=-2  }
\eqn\sclngb{
  [e^{\bar\mu}_\mu]_\epsilon=[e^{\bar\alpha}_\mu]_\epsilon=-{1\over2},
\ [e^{\bar\alpha}_\alpha]_\epsilon=[e^{\bar\mu}_\alpha]_\epsilon=0}
\eqn\sclngc{[B_{\mu\nu}     ]_\epsilon=1,\ 
            [B_{\mu\alpha}  ]_\epsilon=1{1\over2},\ 
            [B_{\alpha\beta}]_\epsilon=2}
\eqn\sclngd{[A_{\alpha\beta\gamma\delta}]_\epsilon=2,\
            [A_{\alpha\beta\gamma\mu   }]_\epsilon=1{1\over2},\ 
            [A_{\alpha\beta\mu\nu      }]_\epsilon=1,\
            [A_{\alpha\mu\nu\rho       }]_\epsilon={1\over2},\ 
            [A_{\mu\nu\rho\eta         }]_\epsilon=0}

We also need to specify how we choose the parameter ${l'}_p$. Since we
have not analyzed the interaction of the pseudo-sugra fields it is not
clear how to choose it precisely but we can get a bound on it value,
which will be useful in what follows. One obvious choice is that
${l'}_p$ be smaller than our cut-off $L_{uv}$, but one can do better.

We are interested in mapping the system of branes parameterized by
$\lambda_i(x)$ to the $x^\alpha-x^\mu$ space. The width in the
$\lambda$ direction is $\sqrt{N}\over L_{uv}$. After the map the width
in the $x^\alpha$ direction is ${l'}_p^2\sqrt{N}\over L_{uv}$. This is
the region in $x^\alpha$ space that is relevant for our discussion. We
would like that the effect of the 10D equations of motion as we
transverse from one side of the $x^\alpha$ plane to another is such
that (in Minkowski space) will not be bigger than the cut-off
$L_{uv}$. Otherwise our theory is not ${\cal N}=4$ SYM even below the
cut-off. The condition for that can be derived from the spread of a
massless particle with a source at one side of this space. We would
like that the spread of the propagator when it reaches the other side
of the $\lambda$ space will be small enough. The result is that the
following condition has to be satisfied
\eqn\conlp{ {{l'}_p^2\sqrt{N}\over L_{uv}}<L_{uv}.}

\subsec{Coupling the D3-branes to the Pseudo-gravity}

The purpose of the contraction was to generate the 10D pseudo-gravity
which has a chance of exactly reproducing the 1-loop corrected
effective action, and which will become usual sugra on on $AdS_5$
after an appropriate transformation (at least for some of the fields).
There are two elements to this which are how the pseudo-gravity fields
couple to the D3-brane, and what are their new equations of motion in
the $\lambda-x$ space. In this section we will discuss the former. We
will also restrict out attention to a set of simple couplings, some of
which will play a role later on, and only at the linearized level.
Our starting point before the contraction are the couplings of the
standard D3 brane to the standard IIB supergravity. We would then like
to trace what couplings remain when we take $\epsilon\rightarrow
0$. 

Before we proceed to the computations, it is worth mentioning that
although we are deriving the correspondence between operators and
fields in the bulk from a variant of the D3-brane action, the
procedure outlined here is different than in \das. In our case we have
a configuration of branes filling an entire 10 dimensional space
(which is not $AdS$) rather than $N$ branes on top of each other at a
point in $AdS$.

\smallskip

1. Coupling of the Axion-Dilaton pair

The (linear) couplings of the axion-dilaton pair to the D3-brane is
given by
$$\int d^4x (\phi F_+^2 +{\bar\phi} F_-^2)$$ 
where in Minkowski space ${\bar\phi}=\phi^*$. 
When going to the gas picture the coupling becomes
\eqn\axdlcp{\int d^4x d^6\lambda \rho(\lambda,x)
(\phi F_+^2 +{\bar\phi} F_-^2).}  This coupling does not change as
$\epsilon\rightarrow 0$.

\smallskip

2. The 2-form fields

The terms in the couplings of the D3-brane to gravity that are linear
in the $B$ fields (with polarization parallel to the brane) are
\eqn\dbtwoa{\int d^4x 
k^{1\over2}\biggl(F_{\mu\nu} B^{NSNS}_{\mu_1\nu_1}
g^{\mu\mu_1}g^{\nu\nu_1}\sqrt{g}+ 
k^{1\over2}\epsilon^{\mu_1..\mu_4}F_{\mu_1\mu_2}B^{RR}_{\mu_3\mu_4}+}
$$k^{3\over2}(F^3{g^{\rho\sigma}}^3)_{\mu\nu}B^{NSNS}_{\mu_1\nu_1}
g^{\mu\mu_1}g^{\nu\nu_1}
\sqrt{g}
+...\biggr)$$ 
(Although we were not careful with the contraction of
the indices, we were careful to include all the appearances of the
metric since it has a non-trivial $\epsilon$
dimension. $F^3{g^{\rho\sigma}}^3$ stands for a specific quantity made
out of 3 $F_{\mu\nu}$ and 3 $g$ with upper indices).

The total $\epsilon$-dimension of these terms is 0, which means that they
survive the contraction. We will rewrite the term as
\eqn\dbtwoa{\int d^4x 
k^{1\over2}(F_{\mu\nu} B^+_{\mu_1\nu_1}
g^{\mu\mu_1}g^{\nu\nu_1}\sqrt{g}+ 
k^{3\over2}(F^3{g^{\rho\sigma}}^3)_{\mu\nu}(B^-+B^+)_{\mu_1\nu_1}
g^{\mu\mu_1}g^{\nu\nu_1}+...}
where $B^\pm=B^{NSNS}\pm*_4B^{RR}$, $*_4$ denotes the 4D Hodge $*$.

Next we would like to keep the couplings of $B^+$ to $F$ and of
$B^-$ to $F^3$. The quantities that we would like to keep fixed are the
$\hat B$
$$B^+=k^{-{1\over2}} {\hat B}^+$$
$$B^-=k^{-{3\over2}} {\hat B}^-$$ Since $k$ is a smaller than the
cutoff we would like to take it to 0. This will leave us only the
coupling of ${\hat B}^-$ to $F^3$.
 
\smallskip

Although not necessary in the linearized approximation, it is
interesting to discuss singularities in $\epsilon$ that might arise
when we perform the contraction. We do not have a complete analysis
but we can rule out some simple occurrences of such singularities. The
argument is as follows. If we are interested in the coupling of the
brane to several pseudo-gravity fields then because the pseudo gravity
fields carry spin and mass dimension we are restricted as to how they
can correct existing terms in the action. For example we have a term
of the form $\int d^4x BF$ and we may as whether we can try and add
another $B$ field and write a correction of the form $\int d^4x
B^2F$. The question is then whether such a term will have a
singularity in $\epsilon$. The answer is that such terms will not have
any singularities in $\epsilon$. To show this we build out of each
pseudo-gravity field a quantity that is dimensionless (by multiplying
power of $k$) and carries no Einstein indices (but may carry Lorentz
indices. This is done by contraction with the vielbein). It turns out
the the $\epsilon$ dimension of these composite fields is 0. Hence
there will be no $\epsilon$ singularity when correcting non-singular
terms by higher powers of the pseudo-gravity fields.

Another issue are the couplings between between fields in the bulk and
higher dimension operators on the brane. An example for such terms,
which will require special treatment in the next section, would be the
expansion
$$\sum_k {({l'}_p)}^{4(k-1)} \int d^4x \sqrt{g} {(\partial_\mu
x^\alpha)}^{2k} {(g^{\mu\mu})}^k{(g_{\alpha\alpha})}^k$$ where we have
indicated only schematically the location and type of important
indices. Although it is difficult to rule out systematically all such
terms, we can rule some on a case by case analysis. For example, this
specific term is such that it is actually an expansion in ${l_p}'\over
L_{uv}$. The reasons is that the kinetic term $\partial x^\alpha$ is
actually derived from the field theory quantity $\partial \lambda$ and
is therefore of order $1/L_{uv}^2$ at most, and the expansion is
therefore in ${{l'}_p}^2{\partial\lambda}\sim {{{l'}_p}^2\over
L_{uv}^2}$.

\subsec{Contraction of the Equations of Motion}

The main purpose of the contraction was to obtain fields whos wave
equation is only 6D. In this section we will briefly discuss how that
comes about. We will not discuss the most general equation of motion but
rather assume that $g_{\alpha\mu}=0$. This will be the case that we
will need in the following.

\smallskip

1. The Axion-Dilaton pair.

The equation of motion of the Axion-Dilaton pair was in fact used
before to motivate the construction. Let us briefly repeat the
argument. The equation of motion, in the linearized approximation, was
\eqn\dileq{ {1\over\sqrt{g}}
\bigl({\partial\over\partial x^\alpha} g^{\alpha\beta}\sqrt{g}
 {\partial\over\partial x^\beta }+ {\partial\over\partial x^\mu }
 g^{\mu\nu} \sqrt{g} {\partial\over\partial x^\nu
 }\bigr)\Phi(\lambda,x)=0.}  Under the contraction the 2nd term is
 $O(\epsilon)$ compared to the first and therefore we obtain, at the
 linearized level for $\Phi$, that the equation of motion for is
\eqn\cntrctd{ {1\over\sqrt{g}}
\partial_\alpha g^{\alpha\beta} \sqrt{g}\partial_\beta\Phi(\lambda,x)=0}
which is a 6D equation of motion.

\smallskip

2. The 2-form fields

The linearized equation of motion for the 2-form fields in IIB sugra is
\gsw\schwrzb 
\eqn\linb{D^P G_{MNP}=-{2\over3} ikF_{MNPQR}G^{PQR}} where $M,N,P,Q,R$ 
run from 0 to 9\foot{G was described above. It is a combination of the
$dB$ that depends on the scalar fields}. When we divide the coordinate
we obtain that the LHS (neglecting the various $\sqrt{g}$ and
$g^{\alpha\mu}$ which will not change the argument\foot{$\sqrt{g}$
terms appear both in the numerator and the denominators, as in
equation \dileq\ and therefore do not contribute to the
$\epsilon$-scaling. These terms, however, will be important later
where we will treat them more carefully.}) are
$$g^{\mu\nu} D_\mu G_{[\nu\rho\eta]}+ g^{\alpha\beta}D_\alpha
  G_{[\beta\mu\nu]}$$
$$g^{\mu\nu}D_\mu         G_{[\nu\rho\gamma]}+
  g^{\alpha\beta}D_\alpha G_{[\beta\rho\gamma]}$$
$$g^{\mu\nu}D_\mu         G_{[\nu\gamma\delta]}+
  g^{\alpha\beta}D_\alpha G_{[\beta\gamma\delta]}$$
These, however, simplify under the contraction.

Suppose we focus on a certain $G_{MNP}$. The scaling of the B fields
is such that the more $\alpha$ indices they have, the higher is their
$\epsilon$ dimensions. This implies that the only term that remains
under the contraction is the one in which the derivative (in $G\sim
dB$) is such that it is in the $x^\alpha$ directions. Furthermore, the
contraction with $g^{\mu\nu}$ adds another power of $\epsilon$ and
makes the term disappear even faster as $\epsilon\rightarrow 0$. The
final result is that the 2nd order term in the equations of motion are
6D and are qualitatively (up to factors of $\sqrt{g}$ which we will
restore later)

\eqn\fineqb{ g^{\alpha\beta}D_\alpha \partial_\beta B_{[\mu\nu]}}
$$ g^{\alpha\beta}D_\alpha \partial_{[\beta}B_{\gamma]\rho}$$
$$ g^{\alpha\beta}D_\alpha \partial_{[\beta} B_{\gamma\delta]}$$

Similar arguments also show that on the RHS, one obtains only
derivatives with respect to $x^\alpha$, so the overall result is that
these equations are also 6D.

\smallskip

3. The RR self-dual 4-form

Since the equations of motion of the 4-form field are of slightly
different nature, it is worthwhile to check them as well. The same
mechanism that helped us in contracting the equations of motion for
the 2-form field strength is again at work for the self-dual
4-form. The self duality equation are a set of equations for the
components
$$F_{\alpha\mu_1\mu_2\mu_3\mu_4},\
  F_{\alpha_1\alpha_2\mu_1\mu_2\mu_3},\
  F_{\alpha_1\alpha_2\alpha_3\mu_1\mu_2}$$ which relates them to their
  dual. However, each component of F is dominated (as
  $\epsilon\rightarrow 0$) by the allowed component of $A_4$ with the
  least number of $\alpha$ indices. If we denote by $d_6$ the exterior
  derivative using on the 6 $\lambda$ coordinates then this implies
  that
$$F_{\alpha\mu_1\mu_2\mu_3\mu_4}\rightarrow 
d_6 A_{\mu_1\mu_2\mu_3\mu_4}$$ 
and its dual satisfies
$$F_{\alpha_1\alpha_2\alpha_3\alpha_5\alpha_5}\rightarrow 
d_6 A_{\alpha_1\alpha_2\alpha_3\alpha_4}.$$

If we denote the components of $A_4$ by $A_{p,q}$ where $p$ denotes
the number of $\mu$ indices and $q$ the number of $\alpha$ indices
then the contracted equations of motion are
\eqn\mtnslfd{d_6 A_{4,0}=*d_6 A_{0,4}}
$$       d_6 A_{3,1}=*d_6 A_{1,3}$$
$$       d_6 A_{2,2}=*d_6 A_{2,2}$$
which are again 6 dimensional equations of motion.

\newsec{Transforming back to the AdS}

We have so far obtained a set of fields which, at least for some of
them, after transforming to the AdS would become the supergravity
multiplet on that space. We would now like to see how perturbing the
field theory by a local operator corresponds to turning on a space
dependent field in the sugra multiplet on AdS. The two classes of
operators that we will discuss are those that couple to the
axion-dilaton pair and part of those that couple to the NSNS and RR
2-form fields. We will discuss only rudiments of this map, and there are
clearly many more details to check.

Idealy one would like to show that for any configuration of the branes
in the path integral, turning on a perturbation in the field theory
corresponds to turning on a field on the sugra. In this case the
statement will be true when we coherently integrate over all the
configuration, which is the classical sugra vacuum that we observe at
the end of the day. It is not clear how to show this for an arbitrary
configuration but we will show it, for some of the sugra fields, under
some dynamical assumptions on the configurations that dominate the
path integral.

\subsec{The dynamical assumptions}

Since we have not analyzed the entire non-linear couplings of the
branes to the pseudo-gravity, we are restricted for the most part to
the regime where we can treat the pseudo-sugra fields as small
perturbation. This is possible for all the sugra fields except
$A_{\mu_1\mu_2\mu_3\mu_3},\ g_{\mu_1\mu_2}$ and
$g_{\alpha_1\alpha_2}$, and we will assume that the rest of the fields
are indeed small. For example, in order not to excite a large
$g_{\alpha\mu}$, we are restricted to look at brane configurations
which are almost flat. For $A_{\mu_1\mu_2\mu_3\mu_4}$, $g_{\mu\mu}$
and $g_{\alpha\alpha}$, we can not assume that they are small because
the effect of the branes on them is large.

If we neglect the effect of the bending of the branes on
$g_{\alpha\mu}$, and set it to $0$ at leading order, then the branes
are roughly parallel and we can solve for the back reaction on the
metric and on the self-dual 4-form. Note that we are in better shape
than if we had tried to use the same argument with the standard sugra
multiplet. The reason is that the equations here are 6D, i.e. only in
the $\lambda$ space, so $g_{\mu\alpha}$ is determined only by
$L^\alpha_\mu$ at the same value of $x$ whereas is in the usual sugra
10D equations of motion it would have been determined by the behavior
of the gas of branes at far values of $x$.

Under this assumption one can write down the solution for the metric
and 4-form, which is a continuum version of the 2-cluster solution
described in \juan. One defines a function $f$ which satisfies
\eqn\deff{{\partial\over\partial {x^\alpha}^2} f=N\rho(x^\alpha,x^\mu)} 
(note that it is a 6D equation of motion) and the metric is then given
by \ansatz
\eqn\mtrcf{g_{\mu\nu}=f^{-{1\over2}}\delta_{\mu\nu},
\ g_{\alpha\beta}=f^{1\over2}\delta_{\alpha\beta},\
F_{0123\alpha}=-{1\over4}\partial_\alpha f^{-1}}

We can try and justify the assumption regarding the fluctuation of the
brane in the following way. Since the source for $g_{\mu\alpha}$ is
$L^\alpha_\mu$ we expect that these metric elements will be
proportional to ${l'}_p^2{\partial_\mu\lambda^\alpha}\sim
{{l'}_p^2\over L_{uv}^2}<<1$.

\subsec{The axion-dilaton pair}

We have seen that the coupling of the axion-dilaton pair to the
D3-branes persists after the contraction.  Before we do the
contraction, the action for the axion-dilaton pair and the coupling of
sugra to a density of D3-branes is of the form
\eqn\dilcp{\int d^4x^\mu d^6x^\alpha \sqrt{g} 
g^{ij}\partial_i \phi \partial_j{\bar \phi}+}
$$N\int d^4x^\mu d^6x^\alpha\rho 
\bigl(\phi F_+^2+{\bar\phi} F_-^2\bigr).$$
When we do the contraction, insert the ansatz from the previous
section, and go back to the $\lambda$ coordinates the equation of
motion for $\Phi$ that we obtain is
\eqn\axmtna{\sum_{\alpha=1}^6
{\partial^2\over\partial{\lambda^\alpha}^2} \phi= 
- N\rho F^+_{\mu\nu}F^+_{\mu_1\nu_1}\eta^{\mu\mu_1}\eta^{\nu\nu_1}} 

Let us now perturb the field theory by a chiral operator of the form
\eqn\perbf{\int d^4x \alpha(x) Tr(F_+^2P(\lambda)),} where $P(\lambda)$ 
is a symmetric traceless polynomial on of $SO(6)$.  
In the ``gas of brane'' approximation we are adding to the action in
the $\lambda-x$ space a term
\eqn\pertg{ \int d^4xd^6\lambda\rho(\lambda,x) \alpha(x) 
F_+^2(\lambda,x)P(\lambda).}
The action now (in addition to the tree level action) is 
\eqn\acta{\int d^4xd^6\lambda\rho(\lambda,x) \bigl(\phi+\alpha(x) 
P(\lambda)\bigr)
F_+^2(\lambda,x).} The equation of motion \axmtna\ is not modified.

We can now see how $\Phi(\lambda,x)$ is turned on in
$AdS$. If we define ${\tilde
\phi}(\lambda,x)=\phi(\lambda,x)+\alpha(x)P(\lambda)$, then this field
satisfies \axmtna\ with the same source terms. In fact, to this order
$\tilde \phi$ appears in the same way as that $\phi$ appeared in the
system before the perturbation. The reason for this is that the
tracelessness condition on $P(\lambda)$ is equivalent to the statement
that
$${\partial^2\over\partial\lambda^2}\alpha(x)P(\lambda)=0.$$ This
implies that whatever field we have on the AdS (which should give
$<\phi>=0$ in the vacuum) now changes by the image of
$\alpha(x)P(\lambda)$ under the transform. Since this function
satisfies ``Laplace equation'' in the $\lambda-x$ plane, it will
clearly satisfy Laplace equation, which is the equation of motion for
this field on AdS.

Using arguments as in section 3.4 one see that the solution is exactly
as describes in
\wittena\igor. Shifting the value of a scalar field on the $\lambda$ 
space by $P(\lambda)\delta(x)$ exactly corresponds to turning on the 
correct boundary-bulk propagator in the $AdS$ bulk.

\subsec{The RR and NSNS B fields}

One can repeat the analysis for the NSNS and RR 2 form
fields. Combinations of these fields couple to $F$ and to $F^3$ on the
D3 brane. We have also analyzed part of their equations of motion on the
$\lambda-x$ plane. Let us focus on the couplings which are
\eqn\cplbfldc{\int d^4x d^6\lambda \rho {B^+}^{\mu\nu}F_{\mu\nu},} 
appended by the equation of motion
\eqn\bsrcc{f^{-{3\over2}}{\partial^2\over{\lambda^\alpha}^2} 
B^+_{\mu\nu}=source\ terms} which is what we obtain when we use ansatz
\mtrcf\ in the contracted equations of motion.  We again add a
perturbation of the form
\eqn\bpert{\int d^4xd^6\lambda \rho C^{\mu\nu}(x)F_{\mu\nu}P(\lambda)} 
and, as before, can reabsorb this term by a shift of $B$ that is
compatible with the equation of motion if $P(\lambda)$ is a symmetric
traceless polynomial. The spectrum of eigenvalues of the AdS Laplacian is
$k(k+4),\ k>1$ and it corresponds the operators $F_{\mu\nu}P(\lambda)$
which agrees with the AdS analysis \foot{The case K=1 corresponds to
adding a perturbation in the decoupled U(1).}.

\newsec{Discussion}

In this paper we started from an effective description of the field
theory and constructed fields that seem to have properties of sugra
fields on $AdS_5\times S^5$. We then showed how one can use to this
construction to understand some aspects of the matching between
operators on the boundary and fields in the bulk. In particular one
sees that perturbing the field theory by a local operator corresponds
immediately to turning on a field on $AdS$.

There are several caveats to this construction. The first is that in
order for it to work a large number of details have to work, for
example, the matching of all operators. Most of these will have to
await future investigation, although we expect that many will work due
to supersymmetry and the fact that we have established them for at
least one field in the multiplet. Another serious problem is that we
were able to calculate the transform only for linear fluctuations
around the $AdS$, but already in our analysis we required non-linear
analysis of the pseudo-sugra multiplet when we analyzed the back
reaction of the metric and 4-form field strength. Extending the
transformation to strong field strengths is a prerequisite before
discussing the full non-linear dynamics of sugra. For example we would
like to calculate 2-pt functions using this prescription. This already
requires analyzing the Lagrangian (which is not a linear functional of
the fields). Another issue might be to try and study non-linear field
theoretic corrections to the interactions of the pseudo-sugra fields
with the branes or with themselves, and map these to $AdS$.

There are also several improvements that one may try and examine. The
most interesting is the following. Even for the case of D3-branes on a
sphere one expects that there is probably a more complete story in
which off-diagonal terms are taken into account. One possible
extension will be the following. In our description we kept the
quantity $F_{\mu\nu}(\lambda,x)$. There could be a picture in which
one keeps also local non-abelian terms, for configurations which are
almost abelian. Such an object would be for example an effective
$[X_\mu,X_\nu](\lambda,x)$ calculated on branes that passes through
the point $\lambda,x$ up to a certain uncertainty, i.e., the matrices
are almost diagonal and this term will measure the deviation from
being diagonal. From the sugra point of view it is natural to have
such terms for the following reason. The brane field $F_{\mu\nu}$ can
probably be thought of as equivalent under a gauge transformation to
$B_{\mu\nu}$ - we are familiar with this when we have a single brane
in 10D spacetime, so when the gas of branes fills spacetime these
might be thought of a one being gauge equivalent to the other
througout spacetime. A non-abelian extension of the local degrees of
freedom, of the form above, might be similarly related to
$B_{\mu\nu}$. 

\centerline{Acknowledgment}

I am indebted to N. Seiberg for collaboration at the initial stages of
this work. I would also like to thank O. Aharony for useful
discussions. This work is supported by grant NSF-PHY 9513835.

\listrefs 

\end